\documentclass[aps,epsf,preprint]{revtex4-1}
\usepackage{amsmath,amssymb}
\usepackage{color}
\def\cred{\color{red}}

\def\1{{\bf 1}}
\def\[{\left[}
\def\]{\right]}
\def\be{\begin{eqnarray}}
\def\ee{\end{eqnarray}}
\def\bm{\begin{pmatrix}}
\def\em{\end{pmatrix}}
\def\nn{\nonumber}
\def\({\left(}
\def\){\right)}
\def\bk#1{\langle#1\rangle}
\def\eq#1{(\ref{#1})}

\def\a{\alpha}

\def\s{\sigma}
\def\o{\omega}
\def\e{\epsilon}

\def\f{\phi}
\def\q{\psi}

\def\l{\lambda}
\def\m{\mu}
\def\x{\times}

\def\p{\partial}
\def\d{\delta}

\def\labels#1{\label{#1}}
\def\edc{\end{document}}
\def\P{{\cal P}}

\def\bn{\begin{enumerate}}

\def\en{\end{enumerate}}
\def\b{\beta}
\def\g{\gamma}

\def\ba{\begin{array}}
\def\ea{\end{array}}
\def\bc{\begin{center}}
\def\ec{\end{center}}

\def\edoc{\end{document}}

\def\^{$\wedge$}
\def\.{\!\cdot\!}

\def\igw#1{\includegraphics[width=#1cm]}
\def\+{\!+\!}
\def\-{\!-\!}
\def\vs{\vskip.5cm}

\usepackage{graphicx}
\graphicspath{/Users/harrylam/Current/filetex/Helicity/graphics/}
\def\S{Sec.~}
\def\tl{{\tilde\l}}
\def\da{{\dot a}}
\def\db{{\dot b}}
\def\tt{{\tilde t}}
\def\P{{\cal P}}
\def\N{{\cal N}}
\def\bs{\boldsymbol{\s}}
\def\bt{\boldsymbol{\tau}}
\def\O{{\cal O}}
\def\ts{{\tilde s}}
\def\?{{\cred ?}}

\begin{document}
\title{Holographic Scattering Amplitudes}
\author{C.S. Lam$^{1,2,3}$}
\email{Lam@physics.mcgill.ca}
\address{$^1$Department of Physics, McGill University\\
 Montreal, Q.C., Canada H3A 2T8\\
$^2$Department of Physics and Astronomy, University of British Columbia,  Vancouver, BC, Canada V6T 1Z1 \\
$^3$CAS Key Laboratory of Theoretical Physics, Institute of Theoretical Physics, Chinese Academy of
Sciences, 55 Zhong Guan Cun East Road, Beijing 100190, China\\}

\begin{abstract}
Inspired by ancient astronomy, we propose a holographic description of perturbative scattering amplitudes, 
as integrals over a `celestial sphere'. Since Lorentz invariance, local interactions, and particle propagations
all take place in a four-dimensional space-time, it is not  trivial to accommodate them in a lower-dimensional
`celestial sphere'. The details of this task will be discussed step by step, resulting in the Cachazo-He-Yuan
(CHY) and similar scattering amplitudes, thereby providing them with a holographic non-string interpretation.
\end{abstract}

%\pacs{}s
\narrowtext
\maketitle

\section{Introduction}
It is well known that it takes the sum of many Feynman diagrams to produce a scattering amplitude, even in the tree approximation.
The discovery of  the Parke-Taylor formula  \cite{PT}, giving  a one-term expression for any gluon
amplitude with all but two identical helicities, 
prompted much research in the past thirty years to generalize its magic \cite{BCFW} and to
obtain compact formulas for other amplitudes. See  \cite{EH} for a review. In particular, the pioneering work of Witten \cite{eW}
using twistors of Penrose \cite{rP} to interpret amplitudes in a string language has been very influential, giving rise to 
many expressions of scattering amplitudes as integrals over  twistor variables,
 or over  string  world-sheet variables \cite{lit}. 
One of the latest  is the Cachazo-He-Yuan
(CHY) formula for tree amplitudes \cite{CHY1,CHY2,CHY3,CHY4,CHY5}, valid in any number of space-time dimensions.
Its generalization to one-loop amplitudes has also been attempted \cite{loop}.

An ordinary string theory contains multiple string excitations, with one of  the two world-sheet variables describing
 length measured along the string. Elementary particles have neither excited states, nor an internal dimension, 
so it seems odd that the most successful interpretation of the CHY and similar amplitudes to date is via string theories \cite{chystring,GLM}.
One obvious reason is the presence of world-sheet complex variables in these formulas, which  naturally suggests a string interpretation.
In order to avoid it, an alternative explanation for these variables must be found.
Inspired by ancient astronomy, we suggest that the complex plane should be interpreted as a 
Riemann sphere, or  rather a `celestial sphere' to make it more physical.
With that interpretation, the CHY formula becomes a holographic formula, expressing the scattering amplitude 
as an integral over the celestial sphere, rather than over configuration space-time variables as in usual quantum field theories. 
The purpose of this note is to discuss how the CHY and other compact formulas can be arrived at in a step by step manner,
 starting from the requirement that it should be a holographic theory similar to that of ancient astronomy.

In astronomy, a star appears to be located 
at the position where its star ray punctures the imaginary celestial sphere, as shown in Fig.~1. In elementary particle
scattering, the `stars' and `star rays' can be thought of as being the particle sources/detectors and the particle beams, 
respectively, and the `celestial sphere' could be taken
to be a microscopic imaginary sphere enclosing the interaction region. Being microscopic, uncertainty relation has to be taken 
into account, which prevents the puncture position
to be determined geometrically like in astronomy. Instead, they must be determined by the surface analogs of 
Klein-Gordon and Weyl equations of motion (\S IIB).

Even with the puncture positions thus determined, there are still many difficulties to overcome for a successful
holographic description of the scattering amplitude.
First of all, Lorentz invariance must
be implemented on the celestial sphere (\S IIA). Moreover, particles interact at
discrete space-time points, not on a two-dimensional `celestial sphere', and they propagate from one space-time point to another,
not on the `celestial sphere'. How these could be accommodated on the `celestial sphere' will be discussed in the following
sections. On-shell and off-shell tree amplitudes are discussed in \S III, spinor helicity amplitudes in \S IV, and scalar loop amplitudes
in \S V. Unlike the string theory where higher genus Riemann surfaces are required to discuss loop amplitudes, making it
very difficult beyond one loop, in a field theoretical approach one simply needs to fold up off-shell tree amplitudes in an appropriate
manner, whatever the number of loops is. In this way, the off-shell holographic amplitude of \S III which possesses  {\it all the correct propagators}
 can be used to obtain a holographic
loop amplitude for any number of loops.

\section{ancient astronomy, holography, and particle physics}
Astronomy is the world's oldest science. 
Long before people could write, observation of celestial  phenomena was already an important
part of their lives. They knew the correlation between temperature and the seasonal position of the sun, as
well as the relation between the
height of the tide and the phase of motion. From the unchanging pattern 
of fixed stars  which appeared day after day everywhere on earth, 
they could have  discovered rotational invariance, time translational invariance, as well as a certain amount of 
spatial translational invariance. In that sense it is the world's oldest science.

Since naked eyes cannot discern distance to the stars and other celestial objects,  they 
all appear to be painted on a two-dimensional imaginary
celestial  sphere. See Fig.~1. 
Astronomy would thus have remained a science of two spatial dimensions if it were not for the motion of Earth,
bringing along information in the third dimension.
Several hundreds of years ago, people noticed a small seasonal variation of the position of some stars.
These variations were attributed to parallax, and to stellar aberrations. 
Parallax, resulting from the different positions of Earth in different seasons, can be used to measure distance to the stars,
thus providing information in the third dimension. 
Stellar aberration, coming from the different relative velocity between earth and the star
rays in different seasons, shows us how to add velocities that agrees with Galilean invariance. If we were able to 
measure tiny parallax and stellar aberration even for distant stars, then we could have obtained not only 
complete three-dimensional spatial
information from observations on a two-dimensional celestial sphere, but also all the kinematic invariants including Lorentz invariance. 

Even dynamics can be deduced from such two-dimensional observations. Newton's discovery of universal gravitation from Kepler's
laws of planetary motion is such an example.

We will refer to any extra information ({\it e.g.}, third dimension)  hidden in the celestial sphere as `holographic information'. 
A hologram yields a three-dimensional image because holographic information is stored in the interference patterns of the hologram.
Astronomy gives us the correct view of space-time because Earth and planetary motions provide us with holographic
information. 
If particle physics can be described by a holographic theory in two spatial dimensions, 
then the theory must  also contain a sufficient amount
of holographic information to yield the correct kinematics and dynamics in our four dimensional space-time. 

\bc\igw{12}{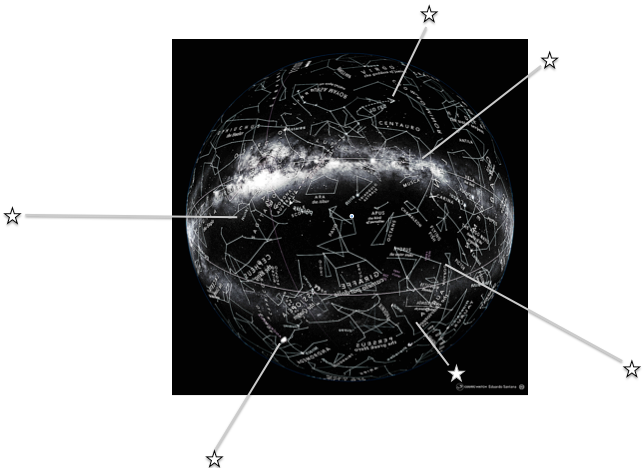}\\ Fig.~1.\quad An imaginary celestial sphere where celestial objects appear to reside.
This picture is also valid for particle scattering, if we interpret the stars and star rays to be particle sources/detectors and
particle beams, respectively, and the celestial sphere to be an imaginary sphere of microscopic size enclosing the interaction region. \ec

There is a similarity between astronomy and  scattering experiments in particle physics
which makes a holographic scattering theory of particle physics somewhat plausible. 
The incoming beams in a scattering experiment are like the star rays, 
with the particle sources  being `stars'.
Detectors and the outgoing particles are also like stars and star rays,  except in reverse. The imaginary `celestial sphere'
could be a tiny sphere enclosing the interaction region, though in astronomy we look outward from inside
the celestial sphere, and in particle physics we look inward from the outside. This distinction however does bring about an important 
difference between the two. In astronomy, the star or planetary position $\s_i$ on the celestial sphere is
just the puncture  of the star ray on the sphere. Its variation with external conditions $E$ (such as time or season) provides the holographic
information. 
With the tiny `celestial sphere' in particle physics,
uncertainty relation prevails. The incoming beam is a plane wave much wider than the interaction region, so it is impossible to fix the punctures 
by geometry.  They must be determined by  a different means,  via a set of {\it scattering equations}, 
which come from  the surface analogs
of  the Klein-Gordon and Weyl equations of motion in field theory. Holographic information is provided by external momenta
and polarizations.

Even with the punctures thus determined, it is still highly non-trivial to be able to express a scattering amplitude $M$ as
a function  of $E$ and  $\s_i$. Particle interactions and propagations take place in four-dimensional space-time, not on a two-dimensional
celestial sphere. 
To have a function on the sphere to describe the scattering, this function must  implicitly contain
 vertices and propagators, in such a way to ensure Lorentz invariance. We shall devote the rest of this paper to discuss, step by step,
 how this can be achieved.  As a start, 
we will review in the next subsection how Lorentz transformation can be implemented on a celestial sphere.

\subsection{Lorentz group representation on a sphere}

By a stereographic projection, a sphere can be mapped onto its equatorial plane. The line joining the north pole and a point A
on the sphere intersects the equatorial plane at a point B, establishing a correspondence A$\leftrightarrow$B between the
sphere and the plane.
The coordinates $(x,y)$ of point B can be represented by a complex number $\s=x+iy$, which will also be used to designate the point 
A as well. No distinction will be made between the sphere and the complex plane in this note, and a
Lorentz transformation on the sphere will simply be specified by the corresponding Lorentz transformation on the 
plane of complex numbers.

The Lorentz group SO(3,1) is locally equivalent to the SL(2,C) group of $2\x2$ complex matrices with determinant 1. Such matrices
are specified by 3 complex numbers, or six real numbers, which describe the three rotations and three boosts of a Lorentz
transformation. More specifically, an SL(2,C) matrix
\be g=\bm \a&\b\cr \g&\d\cr \em\quad (\a\d-\b\g=1)\nn\ee
transforms a complex two-dimensional vector (spinor) $\l=(\l_1,\l_2)^T$  into $\l'=g\l$, and a $2\x2$ complex matrix
\be P=\s_\mu p^\mu=\bm p^0+p^3&p^1-ip^2\cr p^1+ip^2&p^0-p^3\cr \em\nn\ee
into $P'=gPg^T=\s_\mu{p'}^\m$. Since $\det(P)=p^\m p_\m$ and $\det(g)=1$, this transformation preserves the norm of the four-vector
$p^\mu$, showing that SL(2,C) is locally equivalent to the Lorentz group.

The ratio $\s=\l_1/\l_2$ transforms into $\s'=(\a\s+\b)/(\g\s+\d)$. Although the Lorentz group has only a trivial {\it linear} representation
in one dimension, it does have a non-trivial one as shown above when it is represented non-linearly. Hitherto the denominator
$(\g\s+\d)$ will be denoted by $\xi_g(\s)$, or simply $\xi(\s)$.

If $\q$ is a Lorentz spinor or tensor, transforming according to $\q\to G\q$ under a linear Lorentz transformation,
then  $\q(\s)$ will be called a spinor or tensor {\it density of weight $w$} on the sphere if it transforms 
according to $\q(\s)\to G\q(\s)\xi(\s)^w$.
Similarly, $\q(\bs)\equiv\q(\s_1,\s_2,\cdots,\s_n)$ will also be called a spinor or tensor density of weight $w=(w_1,w_2,\cdots,w_n)$ if 
$ \q(\s_1,\s_2,\cdots,\s_n)\to G \q(\s_1,\s_2,\cdots,\s_n)\prod_{i=1}^n\xi(\s_i)^{w_i}$. In particular, if $w_i=\o$ for all $i$, then $\q(\s_1,\s_2,\cdots,
\s_n)$ is said to have a {\it uniform weight} $\o$. The Lorentz-invariant scattering amplitude $M$ will be obtained by assembling
various densities of these types to get a total weight 0.

For later usage, here are some sample weights that can be obtained by a straight forward calculation. 
$1/\s_{ij}\equiv 1/(\s_i-\s_j)$ has a weight $w=(1,1)$, and $1/\s_\a\equiv 1/\prod_{i=1}^n\s_{\a_i\a_{i+1}}$ has a uniform weight $\o=2$, where $\a=(\a_1,\a_2,\cdots,\a_n)$ is a permutation of $(1,2,\cdots,n)$
with $\a_{n+1}\equiv\a_1$. The differential $d\s_i$ has a weight $w_i=-2$. 

In the rest of this subsection,  the difference between the present holographic approach 
and that of the AdS$_3$/CFT$_2$
correspondence in the literature
\cite{AdS3} is explored. These discussions have no bearing on the rest of the article so they can be safely skipped. 

The Lorentz group in a (d+1)-dimensional {\it Minkowskian} space-time is SO(d,1). For a theory possessing scaling and conformal
invariance, this symmetry group is enlarged to the conformal group SO(d+1,2). To be conformal, the theory is required not to
carry any dimensional parameter such as mass. 

There is a mathematical  analog in a d-dimensional {\it Euclidean} space, whose 
symmetry group is the rotation group SO(d), and its conformal extension  is the group SO(d+1,1). 
In particular, for the celestial sphere or the complex plane with d=2, its conformal
group SO(3,1) is the Lorentz group in four-dimensional space-time. It is locally equivalent to
SL(2,C),  
the globally-defined conformal group of the complex plane. However, unlike the Minkowskian conformal group, 
it allows dimensional parameters such as mass to be present, 
because the scaling operation in the complex plane, 
$\s\to\s'=\a^2\s$, is just a Lorentz boost along the third spatial dimension, where parameters such as mass are not affected.

For physical clarity we prefer to think of SO(3,1) as a Lorentz group
of space-time rather than a conformal group of the complex plane. This is where we differ from the AdS$_3$/CFT$_2$
approach; anti-deSitter spaces such as AdS$_3$ never enters into our discussions.
There is another reason to regard SO(3,1)$\sim$SL(2,C) as a Lorentz group  rather than a conformal group, in spite of the fact that we need Lorentz-group representations on the complex plane.
If we were to consider it as a conformal group, then the natural objects to study would be the conformal fields, which carry only {\it abelian} spin quantum numbers. In contrast, in our discussions,
we need (non-abelian) spinors and vectors in four-dimensional space-time, in the form as Lorentz densities on the sphere.

\subsection{Holographic scattering amplitude}
By an  $n$-particle holographic scattering amplitude, we mean an
amplitude that can be expressed as an integral over the `celestial sphere',
\be M=\int d\s_1 d\s_2\cdots d\s_n A(E,\s_1,\s_2,\cdots,\s_n):=\int d^n\bs A(E,\bs),\labels{1}\ee
where $E$ provides external input such as momentum and polarization  of the scattering particles. The puncture positions $\s_i$ on the sphere are determined by a set of $(n\+x)$ {\it scattering equations}
of the form $\f_i(E,\bs,\bt)=0$, where $\bt=(\tau_1,\cdots,\tau_m)$ are extra auxiliary variables that may or
may not be present. In the simplest case to be discussed in the next section, the auxiliary variables are absent, and $x=0$, so there
are just enough scattering equations to determine all the puncture positions $\s_i$. In general,
we may use a set of $\d$-functions to implement the scattering equation constraints, so that
\be A(E,\bs)=\int d^m\bt\[\prod_{i=1}^{n+x}\d\(\f_i(E,\bs,\bt)\)\]B(E,\bs,\bt),\labels{2}\ee
where $B$ contains the dynamics in such a way that $M$ remains Lorentz invariant.

The form and the number of scattering equations $\f_i=0$ depend on  whether the external inputs $E$ are just the momenta $k_i$
of the scattering particles, or momenta $k_i$ plus polarizations $\e_i$
in the spinor-helicity form. We will study  these two cases separately. 

\section{momentum input alone}
In that case, suppose $\s_i$ is the puncture made by the incoming beam with momentum $k_i$ on the `celestial sphere'. 
Then it is convenient to
construct a vector density  $k(\s)=\sum_{i=1}^nk_i/(\s-\s_i)$ to summarize these inputs. This density
 vanishes in the absence of external momenta, 
satisfies the source equation $\bar\p k(\s)=2\pi i\sum_{i=1}^nk_i\d^2(\s-\s_i)$, and is a vector density of weight 2 
as shown below, provided momentum is conserved. Under a Lorentz transformation discussed in \S IIA, it transforms as
\be
k(\s)\to\sum_{i=1}^n{k_i\over\s-\s_i}\xi(\s)\xi(\s_i)=\sum_{i=1}^n{k_i\over\s-\s_i}\xi(\s)\[\xi(\s)-\g(\s-\s_i)\]=\xi(\s)^2k(\s).\labels{3}\ee
The second term within the square bracket vanishes on account of momentum conservation. Hence it has weight 2.

\subsection{Massless momenta}

If all the incoming
momenta are massless, $k_i^2=0$, 
we also require $k(\s)^2=0$ to be true for all $\s\not=\s_i$. This
requirement on the sphere is the counterpart of  the Klein-Gordon equation $\p^2\f(x)=0$ 
for a massless field in space-time. With
\be k(\s)\.k(\s)=\sum_{i\not=j}{k_i\.k_j\over(\s-\s_i)(\s-\s_j)}=\sum_i{1\over\s-\s_i}\sum_{j\not=i}{2k_i\.k_j\over\s_i-\s_j}=0\nn\ee
for all $\s$, it implies 
\be f_i(\bs)\equiv\sum_{j\not=i}{2k_i\.k_j\over\s_{ij}}=0,\quad {\rm for\ } 1\le i\le n,\labels{4}\ee
which is the CHY scattering equation \cite{CHY1}.
By a calculation similar to \eq{3}, one can also show that $f_i(\bs)$ is a scalar density of weight $w_k=2\d_{ki}$ if momentum is 
conserved.

Unlike
astronomy, where each star ray fixes a single puncture, here each set of initial momenta gives rise to $(n\-3)!$
sets of puncture positions because that is how many solutions the CHY scattering equations yield \cite{CHY1}. 
The $\d$-function in \eq{2} implies a sum over all these sets of positions.
In astronomy, it is the earth's motion that provides holographic
information about the third spatial dimension. Here, it is the values of $k_i\.\ k_j$ that provides the holographic information, not only
for a third spatial dimension, but in principle for any number of extra spatial (and temporal) dimensions.

\subsection{Off-shell and/or massive momenta}
Eq.~\eq{4}  needs to be modified for massive particles, and/or  off-shell external lines. This can be accomplished by
adding a term in the numerator of the scattering equation to modify it to
\be  \hat f_i(\bs)\equiv\sum_{j\not=i}{2k_i\.k_j+\m_{ij}\over\s_{ij}}=0,\labels{5}\ee
with a suitably chosen set of parameters $\m_{ij}=\m_{ji}\ (j\not=i)$.

In order to have a smooth transition back to the massless on-shell limit, 
and in order for the off-shell amplitude to be Lorentz invariant, 
we need to keep $\hat f_i(\bs)$ 
a scalar density with the same weight $w_k=2\d_{ki}$. This requires  $\sum_{j\not=i}(2k_i\.k_j+\m_{ij})=0$, or equivalently,
\be \sum_{j\not=i}\m_{ij}=2k_i^2, \quad 1\le i\le n, \labels{6}\ee
where $k_i^2$ is the off-shell amount of the external momentum $k_i$. There are many possible solutions
to this requirement, as there are $n(n\-1)/2$ unknowns $\m_{ij}\ (i\not=j)$ and only $n$ constraints from the weight requirement.
For example, if  $k_+^2=k_-^2=q^2$ and all other $k_i^2=0$, then one solution of \eq{6} is $\m_{+-}=2q^2$ with all
other $\m_{ij}=0$ \cite{sN1}. Another possibility is $\m_{ij}=-2\kappa_i\.\kappa_j$, where $\kappa_i$ are $d$-dimensional 
vectors satisfying momentum conservation such that $\kappa_i^2=k_i^2$ \cite{sN2, bF}.

However, if we also demand the resulting amplitude to be
the same as that given by the sum of {\it all} off-shell planar Feynman tree diagrams,  then
the solution of $\m_{ij}=\m_{ji}\ (i\not=j)$ for particles of mass $m\ge 0$ is uniquely given by \cite{LY1}
\be \m_{i,i\pm1}&=&k_i^2+k_{i\pm1}^2-m^2,\labels{7}\\
\m_{i,i\pm2}&=&-k_{i\pm1}^2+m^2,\labels{8}\\
\m_{i,i\pm p}&=&0,\quad ({\rm if\ }2<p\le n).\labels{9}\ee
These equations should be interpreted in the following way. Being a planar diagram, the momentum carried by
every propagator is equal to the sum of a set of consecutive external lines. 
Plus signs in the subscripts correspond to clockwise counting, mod $n$, and minus signs correspond to counter-clockwise counting. 
Eq.~\eq{7} applies to two neighbouring lines, 
\eq{8} to two next neighbouring lines, and \eq{9} to two lines with a gap of 2 or more.

For a given pair $(i,j)$, we get different results  $\m_{ij}$ from \eq{7} to \eq{9}
depending on whether we reach $j$ from $i$ clockwise or counter-clockwise. The true answer should be the sum of the two.
This completes the explanation of Eqs.~\eq{7} to \eq{9}.

These equations are derived using the fact that if $S$ is a contiguous set of neighbouring lines, then \cite{LY1}
\be \sum_{i,j\in S, i<j}\m_{ij}=\sum_{i\in S}k_i^2-m^2\labels{10}\ee
must be satisfied in order for a propagator $1/\((\sum_{i\in S}k_i)^2-m^2\)$ to be contained in the amplitude. If the holographic
amplitude contains {\it all} Feynman diagrams, then a propagator is present in the amplitude for {\it every}
consecutive set $S$. In this way one arrives at \eq{7} to \eq{9} after some algebra.

If the amplitude corresponds to a {\it single} Feynman diagram, but not the sum of {\it all} of them, then \eq{10} needs
to be satisfied only for those sets $S$ which give rise to a propagator in that particular Feynman diagram, but not all
possible sets $S$. This calls for many fewer conditions and the solution for $\m_{ij}$ is no longer unique. While \eq{7} to \eq{9} always give a valid solution,
there are other ones as well.

There is an exception when $n=4$. This is so because a
Feynman tree diagram contains $(n\-3)$ propagators. The momentum carried by a propagator,
up to a sign, is equal to the sum of all the external momenta on one side of the propagator, and is also equal to the sum of all the 
external momenta on the other side of
 the propagator. In order for this propagator to be present, there are {\it two} sets of condition \eq{10}
to be satisfied, one on either side of the propagator. 
With $(n\-3)$ propagators, there are $2(n\-3)$ requirements. On top of these, 
$\m_{ij}$ must also satisfy \eq{6} in order for $\hat f_i$ to have the correct weights,
so altogether, there are $2(n\-3)+n=3n-6$ conditions to be obeyed by $n(n\-2)/2$ unknowns $\m_{ij}$. For $n=4$, the two numbers
are equal, which means that the set of $\m_{ij}$ giving rise to a {\it single} Feynman diagram is unique, and therefore it must
be identical to those in \eq{9} to \eq{10}.  Already for $n=5$, there are 10 independent $\m_{ij}$ but only 9
conditions, so the solution is no longer unique.

\bc\igw{8}{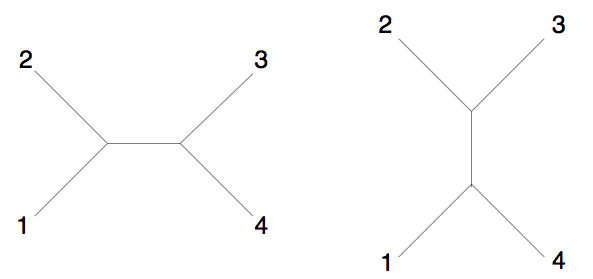}\\ Fig.~2.\quad The $s$-channel (left) and the $t$-channel (right) contributions to a four-point 
planar amplitude with cyclic order (1234) for its external lines. \ec

As an illustration, let us use \eq{10} to obtain directly the unique solution for $n=4$ from {\it one} Feynman diagram.
According to \eq{10}, to produce the propagator in the $s$-channel diagram
in Fig.~2, we need to have
\be \m_{12}=k_1^2+k_2^2-m^2,\quad \m_{34}=k_3^2+k_4^2-m^2.\labels{11}\ee
Together with \eq{6}, these six equations can be used to solve for the six $\m_{ij}$, yielding, other than \eq{11}, also
\be  \m_{13}&=&-k_2^2-k_4^2+2m^2,\quad \m_{14}=k_1^2+k_4^2-m^2,\nn\\
\m_{23}&=&k_2^2+k_3^2-m^2,\quad \m_{24}=-k_3^2-k_1^2+2m^2.\labels{12}\ee
They agree with the rules given in \eq{7} to \eq{9}, and they also automatically contain the $t$-channel propagator
condition in the second and the third equations of \eq{12}.

\subsection{Holographic tree amplitude}
With momentum input alone, the on-shell ($m=0,\ k_i^2=0$) scattering equation $\f_i=f_i(E,\bs)$
contains no auxiliary variable $\bt$, so there are no $\bt$-integrations in \eq{2}. Moreover,
via a suitable Lorentz transformation (which contains three arbitrary complex numbers), one can fix any
three punctures $\s_p, \s_q, \s_r$ to take on any value. In this way the number of $\bs$-integrations can be reduced 
from $n$
to $n\-3$. 
Correspondingly, it can be shown (see the remark below eq.~\eq{15}) that only $(n\-3)$
$f_i$'s are linearly independent, so they can be used to determine the remaining $(n\-3)$ puncture  positions $\s_i$. 
As a result, \eq{1}, \eq{2} can now be replaced by
\be M_n=\int \(\prod_{i\not=p,q,r}d\s_i\d(f_i)\)\s_{pqr}^2 C_n(E,\s_1,\s_2,\cdots,\s_n):=\int d\overline\Omega_{pqr} C_n(E,\bs).\labels{13}\ee
The weight of each $\d(f_i)$ and each $d\s_i$ is $-2$ when $i\not=p,q,r$, so for convenience the quantity $\s_{pqr}:=\s_{pq}\s_{qr}\s_{rp}$ 
has been introduced  to give $d\overline\Omega_{pqr}$ a uniform weight of $\o=-4$.
As a result,
$M$ is Lorentz invariant if $C_n(E,\bs)$ has a uniform weight $\o=+4$. Different choice of $C_n$ corresponds to different 
dynamics, but it turns out that its $\s$-dependence always  consists of products of $1/\s_{ij}$'s as these are the 
fundamental $\s$-quantities containing fixed weights. 

Because of that it is often  useful to convert \eq{13} into a complex-integration form
\be M_n=\int_{\cal O} \(\prod_{i\not=p,q,r}d\s_i{1\over (2\pi i)f_i}\)\s_{pqr}^2 C_n(E,\s_1,\s_2,\cdots,\s_n):=\int_\O d\Omega_{pqr} C_n(E,\bs),\labels{14}\ee
with $\O$ being the contour surrounding every $f_i=0$ counter-clockwise. This form is useful because zeros of $f_i$ 
needed to evaluate \eq{13} are determined by polynomials
of degree $(n\-3)!$ \cite{CHY1}, which are difficult to obtain beyond $n=4$.
In contrast, in the form of \eq{14}, one can distort the contour away from  $f_i=0$ to enclose
poles of $C_n$ in $\s_i$, to allow residue calculus to be used on these explicit poles to evaluate $M_n$.

\subsection{Propagator}
Whatever the dynamics, an $n$-point  tree diagram with cubic vertices
 contains $(n\-3)$ propagators, which  is also the number of $f_i$'s in \eq{14}. That
is not an accident, because it turns out that each integration turns a $1/f_i$  into a propagator.
In this way  the $(n\-3)$ propagators in the Feynman diagram are built up
from the $(n\-3)$ $f_i$'s present, through the  $(n\-3)$ integrations in \eq{14}. Which $f_i$ turns into which propagator in which Feynman diagram depends completely
on the poles of $C_n$ around which the integral is evaluated. 

The denominator of a propagator in a planar Feynman diagram is given by $\(k^S\)^2:=\(\sum_{j\in S}k_j\)^2$ for
some consecutive set $S$ of external lines. Since the scattering function $f_i$ depends both on $k_i\.k_l$ as well as $\s_{il}$,
it seems somewhat miraculous that after evaluation at the poles of $C_n$, whatever they are, the $\s_{il}$'s would always take on values
that turn $f_i$ into  $\(k^S\)^2$. This miracle occurs because of the following sum rule.

For every $i\in S$, define a set of {\it partial scattering functions} by
\be f_i^S=\sum_{j\in S, j\not=i}{2k_i\.k_j\over\s_{ij}}.\nn\ee
If $S$ is the set of all lines $A=\{1,2,3,\cdots,n\}$, then
$f_i^S$ is the scattering function $f_i$ in \eq{4}. Otherwise, it consists of  some but not all the terms in $f_i$.

By a straight-forward calculation, it can be shown that  \cite{LY2}
\be \sum_{i\in S}f_i^S&=&0,\nn\\
\sum_{i\in S}f_i^S\s_i&=&\(k^S\)^2:=\(\sum_{i\in S}k_i\)^2,\nn\\
\sum_{i\in S}f_i^S\s_i^2&=&2k^S\.\sum_{i\in S}k_i\s_i.\labels{15}\ee
In particular, if $S=A$, momentum conservation shows that only $(n\-3)$ $f_i$'s are linearly independent, as previously claimed.
Also, if every $f_i^S=0$ except $i=t$ and $p$, then the first two sum rules imply
\be f_t^S=-f_p^S={\(k^S\)^2\over\s_{tp}}.\labels{16}\ee
It is through \eq{16} that $f_t$ morphs into the inverse propagator $(k^S)^2$, in a way outlined in the following sketch.
For a more detailed explanation please see \cite{LY2}.

We will call an external line $y$ `non-integrating' if the factor $1/f_y$ is absent in the integral \eq{14}.
Initially the constant lines $p, q, r$ are the non-integrating lines.
Let $S$ be a set of external lines containing one and only one  non-integrating line $p$,  
and $m$ other lines $i$,
in such a way that in the limit $\s_{ip}=O(\e)\to 0\  \forall i\in S$,\ $C_n$ behaves like $1/\e^{2m}$.
Now pick any line $t\not=p$
within the set, distort the contour $\O$ in \eq{14} away from $f_t=0$ to enclose the $\e$-pole of $C_n$, but
keeping  it surrounding the rest of the zeros of $f_i$ as before. With $C_n\sim 1/\e^{2m}$, the integrand of \eq{14} would contain a simple
pole in $\e$. The integration $\int d\e$ around $\e=0$ would factorize $M$ into two parts, one containing the lines in $S$,
and the other the lines in its complement $\bar S$. With the help of \eq{16}, $1/f_t$ would turn into the propagator $1/(k^S)^2$, 
linking these two parts.

We can repeat this procedure to expose more propagators in $S$ and   $\bar S$. Each of these two sets contains two non-integrating lines
to choose from, $q,r$ in $\bar S$ and $p,t$ in $S$. Line $t$ has now become a non-integrating line because $1/f_t$ has been morphed
away to become a propagator so it is no longer present. Note also that after the  integration,
every $\s_{ij}$ for $i\in S$ and $j\in\bar S$ becomes $\s_{pj}$ because every $\s_{ip}=O(\e)\to 0$.

This procedure can
be repeated over and over again until all the $(n\-3)$ propagators are exposed. There are many ways of doing it depending on
the order the different propagators are exposed, but the end result yields the same Feynman diagram.

Propagators reflect the time-energy uncertainty condition.
In ordinary quantum field theories, (the denominator of) a propagator $(1/\p^2)\d^4(x)$ emerges
 from the Klein-Gordon equation of motion $\p^2\f(x)=0$ 
when a particle goes off-shell, with the factor $\d^4(x)$ coming directly from canonical quantization.
It is the same propagator whatever the interactions are. Things are very similar in
a holographic scattering theory. The propagator $1/f_t$ comes from the
scattering equation $f_t=0$ when the contour is distorted away from this `on-shell' value
to enclose the poles of $C_n$. Again its presence is 
independent of the dynamics. This suggests that quantization is somehow related to this contour manipulation of
the holomorphic scattering function, though it is not yet clear  in exactly what way.

So far we have concentrated on the on-shell massless amplitudes. For $m\not=0$ and/or $k_i^2\not=0$, correct propagators
will also emerge in the same way because it is this requirement that determines $\m_{ij}$  in eq.~\eq{5}.
The expressions \eq{13} and \eq{14} for the scattering amplitude also remain the same if $f_i$ is replaced by $\hat f_i$.

\subsection{Dynamics}
In  perturbative quantum field theories, dynamics is specified by the vertex. This is also the case in a holographic scattering theory.

The $n$-point holographic amplitude given by \eq{14} contains only $(n\-3)$ integrations. In particular, for $n=3$, 
there is no integration at all so
 the vertex is given simply by $M_3=\s_{pqr}^2C_3(E,\s_p,\s_q,\s_r)$. This is why it is natural to have cubic interactions in
holographic theories. To some extent it is simply a consequence of demanding Lorentz invariance on the `celestial sphere'.

In what follows we shall concentrate on the scalar $\f^3$ theory
 and the pure Yang-Mills theory, though other theories  can be similarly analyzed.

 The vertex of a $\f^3$ theory is simply the coupling constant. For simplicity we shall take it to be 1, hence 
 $C_3(E,\s_1,\s_2,\s_3)=1/\s_{123}^2$.
Similarly, other than the color factor, $C_3$ for the Yang-Mills theory can be obtained from the triple gluon vertex to be
$ C_3(E,\s_1,\s_2,\s_3)=\[(\e_1\.\e_2)(\e_3\.k_1)+(\e_2\.\e_3)(\e_1\.k_2)+(\e_3\.\e_1)(\e_2\.k_3)\]/\s_{123}^2$.
In both cases $C_3$ 
has a uniform weight of $\o=+4$ which renders the scattering amplitude $M$ Lorentz invariant.

Locality in  usual quantum field theory is implemented by demanding contact interactions in space-time. In a holographic
theory without an explicit third spatial dimension, this requirement is replaced by the absence of
a form factor in the three-point interaction $C_3$. In principle, one can multiply the above results by a function of $k_i\.k_j$
without changing its Lorentz weight, but that would be introducing a form factor corresponding to non-local
interactions. Similar remarks also apply to $C_n$ for $n>3$.

To obtain the holographic tree amplitude with the specified dynamics, $C_n$ must be chosen to yield the correct
Feynman $n$-point tree diagrams. In other words, each non-zero contribution to \eq{1} must contain $n\-3$ propagators joining
the proper vertices. 

Let $\a=(\a_1\a_2\a_3\cdots\a_n)$ be a permutation of $(123\cdots n)$. Then 
\be \s_\a:=\s_{\a_1\a_2}\s_{\a_2\a_3}\cdots\s_{\a_{n\-1}\a_n}\s_{\a_n\a_1}\nn\ee
 has a uniform weight  $-2$ for every $\a$. Since $C_n$ must have
a uniform weight of $+4$, the obvious choice in the case of a scalar theory is 
 $C_n=1/(\s_\a\s_\b)$, where $\b$ is another permutation of $123\dots n$ which
may or may not be the same as $\a$. With this choice, $M_n$ is just the color-stripped amplitude of the
CHY bi-adjoint scalar theory \cite{CHY3}. When $\b=\a$, $M_n$ is given by the sum of all planar tree diagrams whose
external lines are ordered according to $\a$. For $\b\not=\a$, only some of these diagrams are summed.

Putting these together, we arrive at the CHY formula for color-stripped scalar amplitude
for massless on-shell particles. In the case $\b=\a=(123\cdots n)$, it is
\be M_n=\int \(\prod_{i\not=p,q,r}d\s_i\d(f_i)\){\s_{pqr}^2 \over\(\s_{12}\s_{23}\cdots\s_{n-1,n}\s_{n1}\)^2}.\labels{17}\ee
It is known that this amplitude is independent of the choice of $p,q,r$ \cite{CHY1,CHY3}. 
For massive and/or off-shell particles, we merely have to replace $f_i$ in \eq{17} by $\hat f_i$ of \eq{5}.

There are other possible choices for $C_n$ that carry the same Lorentz weight $+4$.
For example, we may multiply  $C_n$ of the last paragraph by a function of $k_i\.k_j$. This however introduces a form
factor which makes the theory non-local.  We could also multiply $C_n$ in the last paragraph by a cross ratios
$\s_{ij}\s_{kl}/\s_{ik}\s_{jl}$, or a function of that, but then we will not get the right propagators because the zeros and the poles
present in the cross ratio would ruin the $1/\e^{2m}$ behaviour of $C_n$. This is illustrated in the 
following example.

The two Feynman planar diagrams shown in Fig.~3 are the same,
but drawn differently so that the external lines on the left are ordered according to $\a=(123456789)$,
 and on the right according to $\b=(124395786)$. The amplitude is given by \eq{14} with $C_n=1/(\s_\a\s_\b)$.
 We will first review how this choice of $\a$ and $\b$ in \eq{14} leads to the Feynman diagram in Fig.~3, 
 then we will show how the presence of an additional
 cross ratio ruins it.

The propagators labelled $a$ to $f$ are produced  according to the discussions after \eq{16}. In this illustration, we shall take
the constant lines  to be $p,q,r=2,4,6$, and the propagators to be exposed in the order $a, b, c, d, e, f$.
With the non-integrating lines underlined, and a cap on top of the line $t$ that morphs into a propagator at every step,
the relevant sets $S$  that gives rise to these propagators by having $C_n\sim 1/\e^{2m}$ are
\be S_\a^a&=&\{\hat 1\underline 2\}, S_\b^a=\{\hat 1\underline 2\};\quad S_\a^b=\{\hat 3\underline 4\}, S_\b^b=\{\underline 4\hat 3\};
\quad S_\a^c=\{5\underline 678\hat 9\}, S_\b^c=\{\hat 9578\underline 6\}; \nn\\
 S_\a^d&=&\{\hat 5\underline 678\}, S_\b^d=\{\hat 578\underline 6\};\quad S_\a^e=\{\underline 67\hat 8\}, S_\b^e=\{7\hat 8\underline 6\};
 \quad S_\a^f=\{\hat 7\underline 8\}, S_\b^f=\{\hat 7\underline 8\}.\nn\ee
 
 \bc\igw{6}{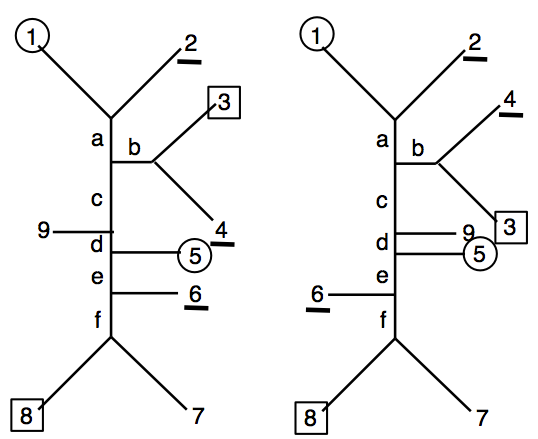}\\ Fig.~3.\quad Two equivalent Feynman diagrams, one showing the $\a$-ordering and the
other the $\b$-ordering. The six propagators are labelled  $a, b, c, d, e, f$,
and the three constant lines $p,q,r$ in \eq{14} are chosen in this illustration to be 2,4,6, shown underlined in the diagrams.
Lines $i,j=1,5$ in the cross ratio $R=\s_{ij}\s_{kl}/\s_{ik}\s_{jl}$ are marked
with a circle, and $k,l=3,8$ are marked with a square. 
\ec
\vs

Now insert the cross ratio $R=(\s_{15}\s_{38})/(\s_{13}\s_{58})$
so that  $C_n=R/(\s_\a\s_\b)$. In Fig.~3, a circle is put
on the lines $i,j=1,5$ and a square is put on the lines $k,l=3,8$. 
How the cross ratio changes after each integration is shown below, using the fact 
that every $\s_{ij}$ for $i\in S$ and $j\in\bar S$ becomes $\s_{pj}$ after the $\e$-integration:
\be
R={\s_{15}\s_{38}\over\s_{13}\s_{58}}\stackrel{a}{\to}{\s_{25}\s_{38}\over\s_{23}\s_{58}}\stackrel{b}{\to}{\s_{25}\s_{48}\over\s_{24}\s_{58}}\stackrel{c}{\to}{\s_{25}\s_{48}\over\s_{24}\s_{58}}\stackrel{d}{\to}{\s_{26}\s_{48}\over\s_{24}\s_{68}}\stackrel{e}{\to}{\s_{26}\s_{46}\over\s_{24}\e}.\ee
When we try to pull out the propagators $a,b,c,d$ successively, $R$ just rides along and changes its value after each integration.
However, when we try to pull out the next propagator $e$, there is an additional $1/\e$ present so that $C_n$ now behaves
like $1/\e^{2m+1}\ (m=2)$ rather than $1/\e^{2m}$. As a result, the $\e$-pole in \eq{14} at this stage becomes a double pole, not a simple
pole anymore, so the result of the integration is something much more complicated than a simple propagator $e$. The introduction of
$R$ into $C_n$ therefore ruins factorization and the end result is no longer a Feynman diagram.

\vs
In the case of Yang-Mills, the generalization of $C_3$ to $C_n$ is the reduced Pfaffian given in \cite{CHY2,CHY3,LY3}. 
The resulting amplitude
is cyclically invariant and gauge invariant, factorizes as in field theory, has no form factor,
and has a uniform weight of $+4$ as required.  Under
these requirements this reduced Pfaffian form is likely to be unique, but I know of no general proof of that. 

\section{Spinor helicity tree amplitude}
With momentum input, the dynamical factor $C_n(E,\bs)$ in a Yang-Mills theory
is proportional to a rather complicated object known as  reduced Pfaffian \cite{CHY2,CHY3}. If  polarization is also added to the input  $E$, then 
 the resulting dynamical factor turns out to be much simpler. 

This is accomplished by using the spinor helicity technique, whose salient feature will be reviewed in the
following subsection. This input changes the scattering equations, and the construction of the scattering amplitude.

\subsection{Spinor helicity technique}
This technique makes use of the local equivalence between SO(3,1) and SU(2)$\x$SU(2),
to represent a light-like Lorentz four-vector $q^\m\in$ SO(3,1)  by a product of two spinors $\l$ and $\tl\in$  SU(2)$\x$SU(2):
\be
(\s_\m q^\m)_{a\da}=q_{a \da}=\bm q^0+q^3&q^1-iq_2\cr q^1+iq^2&q^0-q^3\cr\em=\l_a\tl_\da.\labels{19}\ee
In particular, the momentum $k_i$ of the $i$th massless gluon can be expressed as $\l_i\tl_i$, and a dot product
of two vectors can be written as multiplication of two spinor products,
\be 2k_i\.k_j=(\e_{ab}\l_i^a\l_j^b)(\e_{\db\da}\tl_j^\db\tl_i^\da)\equiv\bk{ij}[ji].\nn\ee
Note that $\e_{ab}=-\e_{ba}$ so $\bk{ij}=-\bk{ji}$ and $\bk{ii}=0$. Similarly $[ij]=-[ji]$.

The polarization vector $\e_i^\pm$ satisfies $\e_i^\pm\.\e_i^\pm=0$ so \eq{19} can again be used to write it as
\be (\e_i^-)_{a\da}&=&\sqrt{2}{\l_i^a\tilde\m^\da\over[\tl\tilde\m]},\nn\\
(\e_i^+)_{a\da}&=&\sqrt{2}{\m^a\tl_i^\da\over\bk{\m\l}},\labels{20}\ee
where the normalization is chosen so that $\e_i^-\.\ \e_i^+=-1$, and gauge dependence
is specified by the arbitrary spinors $\m$ and $\tilde\m$. 
 The spinor-helicity expression for a Lorentz-invariant scattering amplitude $M$ is obtained by making these
substitutions to express it
as spinor products of $\l_i$'s and $\tl_i$'s.

It is important to note that  $k_i^\m$ does not determine $\l_i$ and $\tl_i$ uniquely, for an arbitrary scaling
$\l_i\to s_i\l_i$ and $\tl_i\to\tl_i/s_i\equiv \ts_i\tl_i$  leaves $k_i^\m$ unchanged. Under such a scaling,
$\e_i^-\to s_i^2\e_i^-$ and $\e_i^+\to\ts_i^2\e_i^+$. If $\N$ denotes the set of negative-helicity gluons and $\P$ the set
of positive-helicity lines in a gluon amplitude, the fact that a gluon amplitude $M$ should be linear in each of the polarization vectors
tells us that $M$ should scale like 
\be M\to\(\prod_{n\in\N} s_n^2\prod_{p\in\P}\ts_p^2\)M.\labels{21}\ee 
This relation constrains the number of $\l_i$'s and $\tl_i$'s allowed in the numerator and the denominator of $M$.
 If $m_n$ is
the number of $\l_n$ in the numerator, minus the number in the denominator, plus the number of $\tl_n$
in the denominator, minus that number in the numerator, then according to \eq{21} we should have $m_n=+2$ for every
$n\in \N$. Similarly, if $m_p$ is
the number of $\tl_p$ in the numerator, minus the number in the denominator, plus the number of $\l_p$
in the denominator, minus that number in the numerator, then $m_p=-2$ for every $p\in\P$.
 
There is another constraint coming from the energy dimension of $M$, which should be $4-n$, in the unit when   
the energy dimension of $k$ is taken to be 1 and that of $\l$ and $\tl$ is taken to be ${1\over 2}$. As a result,
the total number of spinor products $\bk{ij}$ and $[ij]$ in the numerator minus those in the denominator should be $4-n$.

These constraints are nicely illustrated in the Parke-Taylor formula for the scattering amplitude 
with $(n\-2)$ positive-helicity
gluons and 2 negative-helicity gluons residing in lines $i$ and $j$ \cite{PT}: 
\be M={\bk{ij}^4\over\bk{12}\bk{23}\cdots\bk{n\-1,n}\bk{n1}}.\ee
The total number of spinor products in the numerator minus those in the denominator is indeed $4-n$. Moreover, $m_n=+2$ and
$m_p=-2$ are also clearly displayed.

\subsection{Spinor scattering equations}
The scattering equations discussed in the last section are constructed via a vector density $k(\s)$ sourced by the input
 momenta $k_i$. In a similar way, when 
both $k_i$ and $\e_i$ are provided as inputs in the spinor helicity form, we could likewise construct  smooth spinor densities
$\l(\s)$ and $\tl(\s)$, sourced by the external spinors, to summarize the input. 
From \eq{20}, we see that negative-helicity gluons provide only for
$\l_n$, not the tilde spinor which is a gauge artifact, and positive-helicity gluons provide only for $\tl_p$, not the un-tilde spinor
which is also a gauge artifact. Furthermore, these spinors are uncertain up to a scaling factor  $t_n$
and $\tt_p$ respectively. Thus the only input that can source a {\it smooth} $\l(\s)$ is $t_n\l_n$, for some suitable scales $t_n$,
and the only input that can source a smooth $\tl(\s)$ is $\tt_p\tl_p$, for some suitable scales $\tt_p$. 
If we double 
the scaling factors in all the sources, $\l(\s)$ and $\tl(\s)$ will remain smooth, but their values must be doubled as well. To
prevent the spinor density $\l(\s)$ to be affected by such a scaling, we should take out from it a smooth scaling factor $t(\s)$,
and similarly a scaling factor $\tt(\s)$ from the spinor density $\tl(\s)$. Hence the relevant densities sourced by the
known spinor helicity inputs should be
\be t(\s)\l(\s)&=&\sum_{n\in\N}{t_n\l_n\over\s-\s_n},\nn\\
\tt(\s)\tl(\s)&=&\sum_{p\in\P}{\tt_p\tl_p\over\s-\s_p}.\labels{23}\ee
Since $k_i=\l_i\tl_i$ for all $i$, it is natural to require $k(\s)=\l(\s)\tl(\s)$ for all $\s$. This requirement, written in the form $\l(\s)k(\s)=\l(\s)(\s_\m k^\m(\s))=0$, is simply the Weyl equation of motion for a spinor. It is the counterpart 
 of the Klein-Gordon equation $k_\m(\s)k^\m(\s)=0$ for scalars on the `celestial sphere'.

With this requirement, it is straight forward to show that \cite{HLW}
the following scattering equations must be satisfied:
\be 
t_p\l_p&=&\sum_{n\in\N}{t_n\l_n\over\s_p-\s_n},\quad (p\in\P)\nn\\
\tt_n\tl_n&=&\sum_{p\in\P}{\tt_p\tl_p\over\s_n-\s_p},\quad (n\in\N).\labels{24}\ee
In other words,
\be  \l_p&=&\l(\s_p),\quad t_p=t(\s_p),\nn\\
\tl_n&=&\tl(\s_n),\quad \tt_n=\tt(\s_n),\ee
which also suggests that we should identify $\tt(\s)$ in \eq{23} with $1/t(\s)$.
These are the scattering
equations used in the ambitwistor string theories \cite{GLM}.

Next, we impose Lorentz covariance. Since $\l_i$ and $\tl_i$
are Lorentz spinors, $\tt_p t_n/\s_{pn}$ in \eq{24} must transform like a Lorentz scalar for every $n\in\N$ and $p\in\P$.
Under a Lorentz transformation, $1/\s_{pn}\to\xi(\s_p)\xi(\s_n)/\s_{pn}$, hence we must have
$t_n\to t_n/\xi(\s_n)$ and  $\tt_p\to\tt_p/\xi(\s_p)$. These complicated transformation laws can be more easily visualized
If we bundle $t_n$ and $\s_n$ into a spinor $\hat\s_n=(\s_n,1)/t_n$,  $\tt_p$ and $\s_p$ into another spinor $\hat\s_p=
(\s_p,1)/\tt_p$, then the fact that
the spinor product $\hat\s_p\.\ \hat\s_n=\s_{pn}/t_n\tt_p\equiv(pn)$ is a Lorentz scalar suggests that 
the spinors $\hat\s_n$ and $\hat\s_p$ are indeed Lorentz spinors. This can be directly verified.

We have now three types of spinor products, $\bk{ij}$ for $\l$, $[ij]$ for $\tl$, and $(pn)=-(np)$ for $\hat\s$. 

With this notation, \eq{23} written in the form
\be
\l(\s)=\sum_{n\in\N}{\l_n\over(\s n)},\ \tl(\s)=\sum_{p\in\P}{\tl_p\over(\s p)},\labels{26}\ee
where $(\s n)=(\s-\s_n)/\tt(\s)t_n,\ (\s p)=(\s-\s_p)/t(\s)\tt_p$, clearly shows that $\l(\s)$ and $\tl(\s)$ are spinor densities of weight 0.
The spinor scattering equations \eq{24}, written in the form, 
\be F_p(\hat\bs)&\equiv&\l_p-\sum_{n\in \N}{\l_n\over(pn)}=0,\nn\\
\tilde F_n(\hat\bs)&\equiv&\tl_n-\sum_{p\in \P}{\tl_p\over(np)}=0,\labels{27}\ee
tells us that the function $\f_i(E,\bs,\bt)$ in \eq{2} should now be identified with $F_p(\hat\bs)$ and $\tilde F_n(\hat\bs)$,
with the auxiliary variables $\bt$ given by $t_n$ and $t_p$. The integration measure in \eq{1} and \eq{2} can now be
combined to be $d^2\hat\s_n=d\s_n dt_n/t_n^3$ and $d^2\hat\s_p=d\s_p d\tt_p/\tt_p^3$.

\subsection{Holographic spinor-helicity amplitude}
Momentum conservation is hidden in the spinor scattering equations. Using \eq{27} and the fact that $(pn)=-(np)$, 
it is easy to show that $\sum_{i=1}^nk_i=\sum_{n\in\N}k_n\+\sum_{p\in P}k_p=\sum_{n\in\N}\l_n\tl_n\+\sum_{p\in P}\l_p\tl_p=0$. The amplitude in \eq{1} and \eq{2} are defined with the momentum-conservation factor $\d^4(\sum_{i=1}^nk_i)$ extracted. This $\d^4$-function is hidden in two $\d^2(F_n)$ or two $\d^2(\tilde F_p)$.
For the sake of definiteness we will take them to be the former from now on, with $n=I, J$, and have them removed before
writing the scattering amplitude $M$. 

With all these considerations, \eq{1} and \eq{2} for the color-stripped gluon amplitude can now be written as
\be
M_n=\int \(\prod_{n\in\N,n\not=I,J}d^2\hat\s_n\d^2(F_n(\hat\bs))\)\(\prod_{p\in\P}d^2\hat\s_p\d^2(\tilde F_p(\hat\bs))\)N_{IJ}D_n,\ee
where $N_{IJ}$ is a normalization factor that depends on $I,J$. 
The energy dimension $4-n$ of the amplitude is now contained completely in the
$\d^2(F_n)$ and $\d^2(\tilde F_p)$, leaving $D_n$ dimensionless. To ensure a local interaction, we shall assume
it to be momentum independent to avoid the appearance of a form factor. To maintain Lorentz invariance, we take
it to be a function of $(ij)=\hat\s_i\.\ \hat\s_j$, invariant under cyclic permutation as is required for a color-stripped amplitude.

To determine $D_n$ and $N_{IJ}$, we resort to the scaling relation \eq{21}. 

First, examine \eq{27}. In order to keep that invariant under a scaling operation \eq{21}, we must also scale every $\hat\s_n$
and $\hat \s_p$ according to
\be \hat\s_n\to s_n\hat\s_n,\quad \hat\s_p\to\ts_p\hat\s_p.\labels{29}\ee
The factor $\bk{IJ}^2\prod_{n\not=I,J}\d^2(F_n)\prod_p\d^2(\tilde F_p)$ then scales exactly like $M$ in \eq{21}.
Under \eq{29}, $(IJ)^2\prod_{n\not=I,J}d^2\hat\s_n\prod_pd^2\hat\s_p$ would also scales like $M$ in \eq{21},
thus if we let $N_{IJ}=\bk{IJ}^2(IJ)^2$, then $D_n$ must scale also exactly like $M$ in \eq{21}, namely,
\be
D_n\to \(\prod_{n\in\N}s_n^2\prod_{p\in\P}\ts_p^2\)D_n.\labels{30}\ee
The easiest way to have a Lorentz-invariant $D_n$ that \eq{30}, cyclic symmetry, as well as factorization is to have
\be
D_n={1\over(12)(23)\cdots(n\-1,n)(n1)}.\ee
As before, adding cyclically permuted cross-ratios like $(ij)(kl)/(ik)(jl)$ would ruin factorization
and therefore not allowed. Putting all these together,
the spinor helicity amplitude for the color-stripped gluon amplitude is given by
\be
M_n=\int \(\prod_{n\in\N,n\not=I,J}d^2\hat\s_n\d^2(F_n(\hat\bs))\)\(\prod_{p\in\P}d^2\hat\s_p\d^2(\tilde F_p(\hat\bs))\){\bk{IJ}^2(IJ)^2\over(12)(23)\cdots(n1)}.\ee
It is known that the amplitude is independent of the choice of $I$ and $J$ \cite{CG}.

\section{Scalar loop amplitudes}
In a quantum field theory of scalar particles, any $\ell$-loop diagram can be obtained  from a tree diagram by folding
$\ell$-pairs of its off-shell lines. If $q_a\ (a=1,\cdots,\ell$) are the momenta of these lines, 
then the loop amplitude is equal to the off-shell tree amplitude, with propagators $1/(q_a^2\-m^2\+i\e)$ inserted
and loop momenta $q_a$ integrated. Since we know how to write
a holographic off-shell tree amplitude {\it with the correct propagators}, this procedure can simply be copied over to obtain a
holographic expression for a scalar amplitude for any number of loops. In the language of \eq{2}, the off-shell momenta $q_a$ 
can be regarded as the auxiliary variables $\bt$.

To illustrate this operation, we shall write down the holographic representation of the 1-loop self energy diagram,
whose usual field-theoretic expression in $4\-\e$ dimension is (Fig.~4)
\be
\Sigma(k)&=&\int{d^{4-\e}q\over(q^2-m^2)((k+q)^2-m^2)}=\int_0^1d\a\int{d^{4-\e}q\over[(1-\a)(q^2-m^2)+\a((k+q)^2-m^2)]^2}\nn\\
&\sim&\int_0^1d\a{1\over\e}\[\a(1-\a)k^2-m^2\]^{-\e},\labels{33}\ee
where $\a$ is the Feynman parameter, and $\sim$ means proportional to. 
\bc\igw{2}{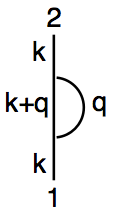}\\ Fig.~4.\quad One-loop self energy\ec

As described above, the holographic 1-loop amplitude is obtained by folding two legs of a 4-point off-shell holographic tree amplitude.
Choosing $p,q,r$ of \eq{14} to be 2,3,4, the 4-point holographic tree amplitude is given by
\be M_4\sim\int {d\s_1\over\hat f_1}{1\over\s_{(1234)}\s_\b},\ee
where $\hat f_1$ is given in \eq{5} and $\m_{ij}$ is shown in \eq{11} and \eq{12}. For the $s$-channel diagram on the left
of Fig.~2, $\s_\b=\s_{(1243)}$, and for the $t$-channel diagram on the right of Fig.~2, $\s_\b=\s_{(1324)}$.

To obtain the loop diagram in Fig.~4, we only
require the $t$-channel diagram of $M_4$, with $k_1=-k_2=k$ and $-k_3=k_4=q$. It is given by
\be
M_4&\sim&\int_{\cal O}{d\s_1\over\hat f_1}{\(\s_{23}\s_{34}\s_{42}\)^2\over(\s_{12}\s_{23}\s_{34}\s_{41})(\s_{13}\s_{32}\s_{24}\s_{41})},
\labels{35}\\ \nn\\
\hat f_1&=&{-m^2\over\s_{12}}+{-2k\.q-k^2-q^2+2m^2\over\s_{13}}+{2k\.q+k^2+q^2-m^2\over\s_{14}}\nn\\
&=&-{m^2\s_{23}\over\s_{12}\s_{13}}-{(2k\.q+k^2+q^2-m^2)\s_{34}\over\s_{13}\s_{14}},\ee
where ${\cal O}$ is a contour surrounding $\hat f_1=0$. To show that this integral does yield the correct $t$-channel diagram given by
$1/\((k+q)^2-m^2\)$, we distort the contour ${\cal O}$ away from $\hat f_1=0$ to surround the poles of $C_4$ instead.
Now $C_4$ contains poles at $\s_{12}=0, \s_{13}=0$ and $\s_{14}=0$, but $\hat f_1$ also contains simple poles at these locations.
The only way for the integrand of \eq{35} to contain a simple pole is to have a double pole present at $C_4$, and this occurs
only at $\s_{14}=0$. The contribution from this pole to the integral is
\be M_4\sim{1\over 2k\.q+k^2+q^2-m^2}={1\over (k+q)^2-m^2},\nn\ee
which is the right result.

Substituting this holographic expression $M_4$ for the $t$-channel propagator into \eq{33}, we get the holographic representation
of the 1-loop self energy amplitude to be 
\be
\Sigma(k)&=&\int{d^{4-\e}q M_4\over(q^2-m^2)}\sim\int {d^{4-\e}q\over(q^2-m^2)}
\int_{\cal O}{d\s_1\over\hat f_1}{\(\s_{23}\s_{34}\s_{42}\)^2\over(\s_{12}\s_{23}\s_{34}\s_{41})(\s_{13}\s_{32}\s_{24}\s_{41})}.\labels{37}
\ee
The result is of course identical to \eq{33}. 

If one wants, \eq{37} can be cast in the form of \eq{1} and \eq{2}, with $n=2$. There are now $m=6$ auxiliary variables,
$\s_3, \s_4$, and the $q^\a$. There are $n+x=4$ scattering equations, $\hat f_1=0$ and three others fixing the values of
$\s_2, \s_3, \s_4$. One can also introduce the Feynman parameter $\a$ to combine the quadratic $q$-dependences in
the denominator, and then carry out the $q$-integration. The result is to replace the $m=6$ auxiliary variables by
$m=3$: $\s_2, \s_3$, and $\a$.

\vs\vs

\begin{acknowledgements}
I am grateful to James Bjorken, Bo Feng, Song He, Yu-tin Huang, Chia-Hsien Shen, and York-Peng Yao for interesting discussions.
\end{acknowledgements}

\end{document}